\documentclass[journal]{IEEEtran}
\usepackage{amsmath}
\usepackage{tikz}
\usepackage{amssymb}
\usepackage{amsfonts}
\usepackage{algorithm}
\usepackage{algorithmicx}
\usepackage{algpseudocode}
\usepackage{dsfont}
\usepackage{float}
\usepackage{cite}
\usepackage{graphicx}
\usepackage{epsfig}
\usepackage{tabularx}
\usepackage{psfrag}
\usepackage{xcolor}
\usepackage{url}
\usepackage[colorlinks,linkcolor=black,urlcolor=black,anchorcolor=black,citecolor=black,hyperfootnotes=true]{hyperref}
\usepackage{bm}
\usepackage{amsthm}
\usepackage{caption}
\usepackage{subcaption}
\usepackage{anyfontsize}
\theoremstyle{plain}

\theoremstyle{definition}
\newtheorem{remark}{Remark}

\def\BibTeX{{\rm B\kern-.05em{\sc i\kern-.025em b}\kern-.08em
    T\kern-.1667em\lower.7ex\hbox{E}\kern-.125emX}}
\usepackage{balance}
\usepackage{lipsum}

\begin{document}

\title{On Channel Estimation for Group-Connected Beyond Diagonal RIS Assisted Multi-User MIMO Communication}
\author{\IEEEauthorblockN{Rui Wang, Junyuan Gao, Shuowen Zhang, Bruno Clerckx, and Liang Liu}
\thanks{Rui Wang, Junyuan Gao, Shuowen Zhang, and Liang Liu are with the Department of Electrical and Electronic Engineering, The Hong Kong Polytechnic University, Hong Kong SAR, China (e-mails: rui-eie.wang@connect.polyu.hk, \{junyuan.gao, shuowen.zhang, liang-eie.liu\}@polyu.edu.hk).}
\thanks{Bruno Clerckx is with the Department of Electrical and Electronic Engineering, Imperial College London, London, U.K. (e-mail: b.clerckx@imperial.ac.uk).}
}
\maketitle

\begin{abstract}
   Beyond diagonal reconfigurable intelligent surface (BD-RIS) architectures offer superior beamforming gain over conventional diagonal RISs. However, the channel estimation overhead is the main hurdle for reaping the above gain in practice. This letter addresses this issue for group-connected BD-RIS aided uplink communication from multiple multi-antenna users to one multi-antenna base station (BS). We first reveal that within each BD-RIS group, the
   cascaded channel associated with one user antenna and one BD-RIS
   element is a scaled version of that associated with any other user antenna and BD-RIS
   element due to the common RIS-BS channel. This insight drastically reduces the dimensionality of the channel estimation problem. Building on this property, we propose an efficient two-phase channel estimation protocol. In the first phase, the reference cascaded channels for all groups are estimated in parallel based on common received signals while determining the scaling coefficients for a single reference antenna. In the second phase, the scaling coefficients for all remaining user antennas are estimated. Numerical results demonstrate that our proposed framework achieves substantially lower estimation error with fewer pilot signals compared to state-of-the-art benchmark schemes.
\end{abstract}

\begin{IEEEkeywords}
Beyond diagonal reconfigurable intelligent surface (BD-RIS), group-connected architecture, channel estimation, low-overhead communication.
\end{IEEEkeywords}

\section{Introduction}

Reconfigurable intelligent surface (RIS) is a key technology for the sixth-generation (6G) networks. Beyond diagonal RIS (BD-RIS) architectures, which use inter-element connections to create non-diagonal scattering matrices, offer more versatile wave manipulation, enhanced beamforming gains, and expanded network coverage over conventional diagonal RISs  \cite{Basar2019wireless,jian2022inte,Li2023BDRIS,shen2022Modeling}. However, channel estimation is a main hurdle to reap the above gains, because the interconnected nature of BD-RIS elements leads to a dramatic increase in the number of channel coefficients needed to be estimated compared to conventional diagonal RIS systems \cite{li2024channel,de2024channel,wang2025lowoverhead}. Several recent studies have explored channel estimation for BD-RIS aided communications. A least squares (LS)-based method was introduced in \cite{li2024channel} to obtain a closed-form estimation of the cascaded channel. The work in \cite{de2024channel} developed two distinct tensor decomposition-based algorithms, utilizing Khatri-Rao factorization (KRF) and alternating least squares (ALS) to reduce estimation overhead. Building on this tensor paradigm, a semi-blind scheme was proposed in \cite{dearaujo2024semiblind}, while the ALS framework was extended to multi-user scenarios in \cite{ginige2024prediction}. A limitation of these methods is that their channel estimation overhead is much higher than that in the conventional RIS aided systems.  

Although the property that cascaded channels of different users share a common RIS-base station  (BS) component has been utilized in conventional RIS systems \cite{wang2020channel,chen2023channel}, this letter focuses on the unique correlation among the cascaded channels associated with interconnected elements in a BD-RIS.
Recently, our work in \cite{wang2025lowoverhead} unveiled a crucial channel property in fully-connected BD-RIS aided systems, i.e., the cascaded channel associated with any user antenna and BD-RIS element is a scaled version of that associated with another user antenna and BD-RIS element, stemming from the
common RIS-BS channel. By exploiting this property, \cite{wang2025lowoverhead} proposed a novel two-phase estimation scheme that first estimates a reference cascaded channel and then only estimates the scaling coefficients of the other cascaded channels, showing that the channel estimation overhead in BD-RIS aided systems is of the same order as that in conventional RIS aided systems \cite{wang2020channel}. 

In this letter, we aim to extend this method to the group-connected BD-RIS aided communication systems. In a group-connected BD-RIS, the reflecting elements are partioned into several disjoint groups and the elements in each group are inter-connected. One heuristic scheme is to turn on one group of elements during each time interval, then the results in \cite{wang2025lowoverhead} can be applied to estimate each group's channels. However, this scheme is not optimal because the channels in each group are estimated individually based on separate pilot signals. Our contribution in this letter is the joint estimation across all groups, which utilizes common pilot signals to determine the reference channels in all groups. This method achieves a further reduction in channel estimation overhead, as verified by numerical results.

\section{System Model}\label{sys_model}

\begin{figure}[t]
    \centering
    \includegraphics[scale=0.1]{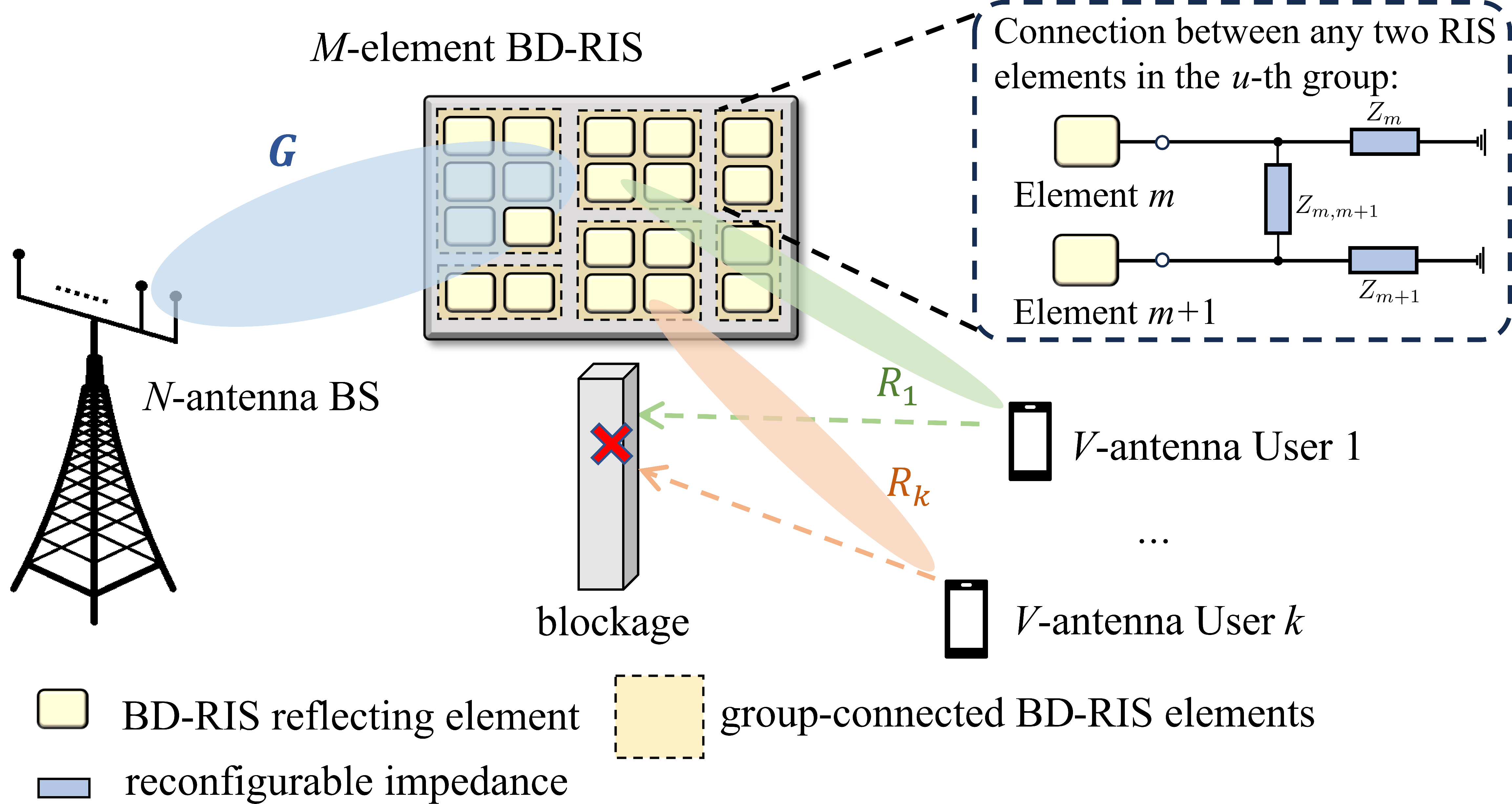}
    \caption{A BD-RIS assisted MU-MIMO uplink communication system.}
    \label{fig:sys_model}\vspace{-17pt}
\end{figure}

We consider an uplink multi-user MIMO communication system. The system, illustrated in Fig. \ref{fig:sys_model}, incorporates a BS equipped with $N$ antennas, a BD-RIS comprising $M$ passive reflecting elements, and $K$ users, each employing $V$ antennas. This letter considers a group-connected BD-RIS architecture. The set of $M$ BD-RIS elements is partitioned into $U$ disjoint groups and the BD-RIS elements in each group are inter-connected. Let $M_u$ denote the number of elements in group $u$, $u=1,\cdots,U$, such that $\sum_{u=1}^{U} M_u = M$. At time instant $t$, the scattering matrix of the BD-RIS, denoted by ${\bm \Phi}_t\in\mathbb{C}^{M\times M}$,  is a block-diagonal matrix
\begin{align}
    &{\bm \Phi}_t = {\rm blkdiag}({\bm \Phi}_{1,t}, {\bm \Phi}_{2,t}, \dots, {\bm \Phi}_{U,t}), ~~\forall t,
\end{align}
where ${\bm \Phi}_{u,t}=[{\bm \phi}_{u,t,1},\cdots,{\bm \phi}_{u,t,M_u}]$, with ${\bm \phi}_{u,t,m}$ denoting the reflecting coefficients of the $m$-th element in group $u$ at time instant $t$, $\forall u$,  $m=1,\cdots,M_u$. According to \cite{shen2022Modeling}, it follows that
\vspace{-5pt}
\begin{align}\label{unitary}
	{\bm \Phi}_{u,t}^H{\bm \Phi}_{u,t}={\bm \Phi}_{u,t}{\bm \Phi}_{u,t}^H={\bm I}_{M_u},~~\forall u,t.
\end{align}

A quasi-static narrowband block fading channel model is considered, under which the channels are assumed to remain approximately constant within each coherence block. In this letter, we assume that the direct links from users to the BS are blocked. The baseband equivalent channel from the $v$-th antenna of user $k$ to the $m$-th BD-RIS reflecting element in group $u$ is denoted by $r_{k,v,u,m}\in\mathbb{C}$, while the channel from the $m$-th BD-RIS element in group $u$ to the BS is given by ${\bm g}_{u,m}\in\mathbb{C}^{N\times 1}$, $k=1,\dots,K$, $v=1,\dots,V$, $u=1,\cdots,U$, and $m=1,\dots,M_u$. The overall channel from user $k$ to the group $u$ of BD-RIS elements is defined as ${\bm R}_{k,u}\in\mathbb{C}^{M_u\times V}$, $\forall k,u$, with the $(m,v)$-th element being $r_{k,v,u,m}$, $m=1\cdots,M_u,v=1,\cdots,V$, and the overall channel from the group $u$ of BD-RIS elements to the BS is ${\bm G}_u=[{\bm g}_{u,1},\dots,{\bm g}_{u,M_u}]\in\mathbb{C}^{N\times M_u}$. Then, the overall channel from user $k$ to the BD-RIS is denoted by ${\bm R}_k=[{\bm R}_{k,1}^T,\cdots,{\bm R}_{k,U}^T]^T\in\mathbb{C}^{M\times V}$, and the overall channel from the BD-RIS to the BS is denoted by ${\bm G}=[{\bm G}_1,\cdots,{\bm G}_U]\in\mathbb{C}^{N\times M}$. Accordingly, the uplink user-RIS-BS cascaded channel from user $k$ to the BS is
\begin{align}\label{H_k}
	{\bm H}_{k,t}&={\bm G}{\bm \Phi}_t{\bm R}_k={\sum}_{u=1}^U{\bm G}_u{\bm \Phi}_{u,t}{\bm R}_{k,u},~~\forall k,t.
\end{align}

It can be shown that at any time instant $t$, we have 
\begin{align}\label{unvec}
	{\bm G}_u{\bm \Phi}_{u,t}{\bm R}_{k,u}={\rm unvec}({\bm J}_{k,u}
	{\bm \phi}_{u,t}),~~u=1,\cdots,U,
	\end{align}
where ${\rm unvec}(\cdot)$ denotes the inverse operation of vectorization, ${\bm \phi}_{u,t}={\rm vec}({\bm \Phi}_{u,t})$ with ${\rm vec}(\cdot)$ denoting the vectorization operation, and
\begin{align}\label{J_ku_1}
	{\bm J}_{k,u}={\bm R}_{k,u}^T\otimes{\bm G}_u,
\end{align}
denotes the cascaded channel associated with user $k$ and the $u$-th group of BD-RIS elements, with $\otimes$ denoting the Kronecker product. Then, the signal received at the BS at time instant $t$ is given as
\begin{align}\label{rev}
    {\bm y}_{t}&={\sum}_{k=1}^K{\bm H}_{k,t}\sqrt{p}{\bm a}_{k,t}+{\bm z}_t \notag\\
    &={\sum}_{k=1}^K{\sum}_{u=1}^U{\rm unvec}({\bm J}_{k,u}
    {\bm \phi}_{u,t})\sqrt{p}{\bm a}_{k,t}+{\bm z}_t, \notag\\
    &\overset{(a)}{=}{\sum}_{k=1}^K{\sum}_{u=1}^U\sqrt{p}({\bm a}_{k,t}^T\otimes{\bm I}_N){\bm J}_{k,u}{\bm \phi}_{u,t}+{\bm z}_t,~~t=1,\cdots,\tau,
    \vspace{-5pt}
\end{align}
where $p$ is the identical transmit power for all users, ${\bm a}_{k,t}=[a_{k,1,t},\cdots,a_{k,V,t}]^T\in\mathbb{C}^{V\times 1}$ is the unit-power pilot signal from user $k$ at time instant $t$, ${\bm z}_t\sim\mathcal{CN}({\bm 0},\sigma^2{\bm I}_N)$ is the additive white Gaussian noise (AWGN) at the BS at time instant $t$, and $\tau$ denotes the length of the pilot sequence in each coherence block, and $(a)$ is because ${\rm vec}({\bm A}{\bm X}{\bm C})=({\bm C}^T\otimes{\bm A}){\rm vec}({\bm X})$ given any matrix $\bm A$, ${\bm X}$ and $\bm C$. In this letter, we mainly focus on the estimation of cascaded channels ${\bm J}_{k,u}$'s based on \eqref{rev}, which is needed for BD-RIS scattering matrix design \cite{li2024channel,de2024channel}, and propose an innovative scheme with low training overhead by exploiting a novel channel property.

\section{Channel Property and Problem Statement}\label{problem}

To estimate ${\bm J}_{k,u}$'s based on \eqref{rev}, existing approaches proposed in \cite{li2024channel,de2024channel} have treated all the $KVN\sum_{u=1}^UM_u^2$ entries in ${\bm J}_{k,u}$'s as independent unknown variables. However, the cascaded channels associated with the BD-RIS elements within each group are highly correlated. Specifically, according to \eqref{J_ku_1}, we have
\begin{align}\label{Jku}
	{\bm J}_{k,u}=
	\left[
	\begin{array}{ccc}
		{\bm Q}_{k,1,u,1} & \cdots & {\bm Q}_{k,1,u,M_u}  \\
		\vdots & \ddots & \vdots \\
		{\bm Q}_{k,V,u,1} & \cdots & {\bm Q}_{k,V,u,M_u}
	\end{array}
	\right]
	\in\mathbb{C}^{VN\times M_u^2},
\end{align}
where
\begin{align}\label{Q_kvum}
	{\bm Q}_{k,v,u,m}=r_{k,v,u,m}{\bm G}_u,~~\forall k,v,u,m=1,\cdots,M_u,
\end{align}
denotes the $(v,m)$-th sub-block of ${\bm J}_{k,u}$. Note that ${\bm Q}_{k,v,u,m}$ can be viewed as the cascaded channel corresponding to the path from the antenna $v$ of user $k$ to the $m$-th element in group $u$ via wireless channel, then to all the elements in this group via inter-connected circuit, and last to the BS via wireless channel. A key observation according to \eqref{Jku} and \eqref{Q_kvum} is that different cascaded channels ${\bm Q}_{k,v,u,m}$'s share a common RIS-BS channel component ${\bm G}_u$, $\forall u$. Due to this common component, according to \eqref{Q_kvum}, we have 
\begin{equation}\label{corr_u}
	{\bm Q}_{k,v,u,m}=\beta_{k,v,u,m}{\bm Q}_{1,1,u,1},~~\forall u,~(k,v,m)\neq(1,1,1),
\end{equation}
where
\begin{equation}\label{beta}
	\beta_{k,v,u,m}=\frac{r_{k,v,u,m}}{r_{1,1,u,1}}.
\end{equation}
This structural insight reveals that the estimation task can be substantially simplified. With each group $u$, we define ${\bm Q}_{1,1,u,1}$ associated with its first element and the first antenna of user $1$ as the reference channel. After the reference channel of each group $u$ is determined,  any other cascaded channels of this group, i.e., ${\bm Q}_{k,v,u,m}$, $\forall (k,v,m)\neq(1,1,1)$, can be reconstructed by merely estimating the scalar $\beta_{k,v,u,m}$. Consequently, the set of independent unknowns comprises the $U$ reference cascaded channels, i.e., ${\bm Q}_{1,1,u,1}$'s, $u=1,\cdots,U$, and $KVM-U$ associated scaling coefficients, i.e., $\beta_{k,v,u,m}$'s, $\forall (k,v,m)\neq(1,1,1)$ and $\forall u$. The total number of independent unknowns is thus $NM + KVM - U$. This number is significantly smaller than the $KVN\sum_{u=1}^{U} M_u^2$ unknowns required by existing schemes such as \cite{li2024channel} and \cite{de2024channel}, which do not exploit the specific channel property in \eqref{corr_u} and \eqref{beta}. In the following, we propose a efficient scheme to estimate the above reduced number of channel coefficients.

\vspace{-5pt}
\section{Proposed Two-Phase Channel Estimation Protocol}\label{protocol}

Define ${\bm B}_{k,u}=[\boldsymbol{\beta}_{k,1,u},\cdots,\boldsymbol{\beta}_{k,V,u}]\in\mathbb{C}^{M_u\times V}$, where ${\bm \beta}_{k,v,u}=[\beta_{k,v,u,1},\cdots,\beta_{k,v,u,M_u}]^T$, and $\beta_{1,1,u,1}=1$, $\forall u$. It can be shown that the received signals given in \eqref{rev} reduce to
\begin{align}\label{y_t_new}
	{\bm y}_t={\sum}_{k=1}^K{\sum}_{u=1}^U\sqrt{p}{\bm Q}_{1,1,u,1}{\bm \Phi}_{u,t}{\bm B}_{k,u}{\bm a}_{k,t}+{\bm z}_t.
\end{align}
In this section, we propose a two-phase channel estimation protocol to estimate all cascaded channels based on \eqref{y_t_new}, which is organized as follows. \textbf{Phase I ($\tau_1$ time instants):} The BS estimates the cascaded channels associated with the $1$st antenna of user $1$, including the reference channels ${\bm Q}_{1,1,u,1}$ for all groups $u=1,\cdots,U$, alongside the scaling coefficients $\beta_{1,1,u,m}$'s for the elements within those groups. \textbf{Phase II ($\tau_2=\tau-\tau_1$ time instants):} The BS estimates the remaining scaling coefficients $\beta_{k,v,u,m}$'s for all other antennas of user $1$ and all antennas of users $k=2,\cdots,K$. Finally, the overall channels ${\bm Q}_{k,v,u,m}$ are recovered based on  \eqref{corr_u}. The detailed implementation of each phase is presented below.

\vspace{-5pt}
\subsection{Phase I: Estimation of the Cascaded Channels Associated with the 1st Antenna of User 1}

In this phase, we aim to estimate the reference cascaded channels ${\bm Q}_{1,1,u,1}$'s for all groups $u=1,\dots,U$, and the scaling coefficients associated with the first antenna of user 1. To implement this, we let only the $1$st antenna of user $1$ transmit non-zero pilot signals while all other
$KV-1$ antennas remain silent, i.e., $a_{k,v,t}=		a_t,$ if $k=1, v=1$, and $a_{k,v,t}=0$ otherwise,  $t=1,\cdots,\tau_1$. Under this condition, the received signal \eqref{y_t_new} simplifies to:
\begin{align}\label{y_t_P1}
	\hspace{-10pt}{\bm y}_t = \sum_{u=1}^U\left({\bm f}_{u,t,1}+\sum_{m=2}^{M_u}\beta_{1,1,u,m}{\bm f}_{u,t,m}\right),~t=1,\cdots,\tau_1.
\end{align}
where ${\bm f}_{u,t,m}=\sqrt{p}a_t{\bm Q}_{1,1,u,1}{\bm \phi}_{u,t,m}$. The primary challenge in estimating $\bm{Q}_{1,1,u,1}$'s and the coefficients $\beta_{1,1,u,m}$'s stems from the fact that the received signals ${\bm y}_t$ in \eqref{y_t_P1} are non-linear functions of them. 
Recently, our work \cite{wang2025lowoverhead} proposed an efficient channel estimation approach in the special case of a fully-connected BD-RIS, i.e., $U=1$. A straightforward method for the case of group-connected BD-RIS is thus as follows. We divide the overall time into $U$ blocks, while at the $u$-th block, we shut down groups $1, \cdots, u-1, u+1,\cdots,U$, and apply the approach in \cite{wang2025lowoverhead} to estimate the channels associated with group $u$, i.e., ${\bm Q}_{1,1,u,1}$ and $\beta_{1,1,u,m}$'s, $u=1,\cdots,U$. However, this approach is sub-optimal because it requires a base received signal and its variant for every group to estimate the associated reference channel, i.e., ${\bm Q}_{1,1,u,1}$, $\forall u$. A unique base received signal is assigned for group $u$ to build a linear function of $\bm{Q}_{1,1,u,1}$ to estimate it, $\forall u$. In practice, this base received signal can be repeatedly used for all groups. Thus, we propose to utilize a common base signal shared by all groups and assign a variant based on it for each group, reducing the overhead to estimate $\bm{Q}_{1,1,u,1}$'s.

Specifically, we divide the overall $\tau_1$ time instants in Phase I into $U+1$ parts to facilitate the differential estimation. The first part provides the base signal, including $\tau_{1,1}$ time instants and the $(1+i)$-th part consisting of $\delta_i\leq \tau_{1,1}$ time instants, $i=1,\cdots,U$, provides the variants for group $i$. The BS-RIS scattering matrices and the user pilot signals in these $U+1$ parts are given as follows: At time instants $t=1,\cdots,\tau_{1,1}$ in the first part, the user pilot signals $a_t$'s are set as random non-zero scalars, and the BD-RIS scattering matrices  ${\bm \Phi}_{u,t}=[{\bm \phi}_{u,t,1},\cdots,{\bm \phi}_{u,t,M_u}]$ are set as random unitary matrices. Let $t_{<i} = \tau_{1,1} + \sum_{j=1}^{i-1} \delta_j$, and $t_{<1}=\tau_{1,1}$, then in the $(1+i)$-th part of Phase I at time instants $t=t_{<i}+1,\cdots,t_{<i}+\delta_{i}$, $i=1,\cdots,U$, the user pilot signal is set as
\begin{equation}\label{pilot_rule}
	a_{t}=a_{t-t_{<i}}.
    \vspace{-3pt}
\end{equation}
Moreover, the BD-RIS scattering matrix ${\bm \Phi}_{u,t}=[{\bm \phi}_{u,t,1},\cdots,{\bm \phi}_{u,t,M_{u}}]$ is set as
\begin{align}\label{Phi_rule}
	\hspace{-10pt}\begin{cases}
	{\bm \phi}_{u,t,m}=e^{{\rm j}\theta}{\bm \phi}_{u,t-t_{<u},m},~\text{if $u=i, m=1$},\\
	{\bm \phi}_{u,t,m}={\bm \phi}_{u,t-t_{<u},m},~~~~\text{otherwise},
	\end{cases}\hspace{-8pt}\forall u,~ m\geq 2,
\end{align}
where j is the imaginary unit, and $\theta\in(0,2\pi)$ is an arbitrary phase shift. Thus, the scattering
matrix at time instants $t=t_{<i}+1,\cdots,t_{<i}+\delta_{i}$ also satisfies \eqref{unitary}.

Given the above strategy, the received signals over $\tau_1=\tau_{1,1}+\sum_{i=1}^{U}\delta_i$ time instants can be re-written as
{\fontsize{9pt}{12pt}
\begin{align}
	&{\bm y}_t=\sum_{u=1}^{U}\left({\bm f}_{u,t,1}+\sum_{m=2}^{M_u}\beta_{1,1,u,m}{\bm f}_{u,t,m}\right)+{\bm z}_t,~~t=1,\cdots,\tau_{1,1}, \label{y_t_base}\\ 
	&{\bm y}_{t_{<i}+t}=\left(e^{{\rm j}\theta}{\bm f}_{i,t,1} +\sum_{m=2}^{M_{i}}\beta_{1,1,i,m}{\bm f}_{i,t,m}\right) +\sum_{u \neq i}\Bigg({\bm f}_{u,t,1} \notag\\
	&+ \sum_{m=2}^{M_{u}}\beta_{1,1,u,m}{\bm f}_{u,t,m}\Bigg) +{\bm z}_{t_{<i}+t},~~i=1,\cdots,U,~t=1,\cdots,\delta_i. \label{y_t_modified}
\end{align}
By subtracting the baseline signals ${\bm y}_t$'s from the variation signals ${\bm y}_{t_{<i}+t}$'s, $t=1,\cdots,\delta_i$, respectively, the contributions from all scaling coefficients and non-target reference channels are eliminated. This yields $\delta_{i}$ effective received signals contributed only by the reference channel of group $i$, ${\bm Q}_{1,1,i,1}$. The $t$-th effective received signal associated with the $u$-th group is
\begin{align}\label{y_bar}
	&\bar{\bm y}_{u,t} = {\bm y}_{t_{<u}+t} - {\bm y}_{t} = \sqrt{p} a_t(e^{{\rm j}\theta} - 1) {\bm Q}_{1,1,u,1} {\bm \phi}_{u,t,1}+{\bm z}_{u,t}, \notag\\
	&u=1\cdots,U,~~t=1,\cdots,\delta_{u},
\end{align}
where ${\bm w}_{u,t}={\bm z}_{t_{<u}+t}-{\bm z}_{t}\sim\mathcal{CN}({\bm 0},2\sigma^2{\bm I}_N)$ is the effective noise. Then, the overall effective received signal of the $u$-th group, i.e., $\bar{\bm Y}_{u,1} = [\bar{\bm y}_{u,1}, \dots, \bar{\bm y}_{u,\delta_{u}}]$, is expressed as
\begin{align}\label{Y1bar}
	\bar{\bm Y}_{u,1} = \sqrt{p}(e^{{\rm j}\theta}-1){\bm Q}_{1,1,u,1} {\bm \Psi}_{u,1}+{\bm W}_u,
\end{align}
where ${\bm \Psi}_{u,1} = [a_1{\bm \phi}_{u,1,1}, \dots, a_{\delta_u}{\bm \phi}_{u,\delta_{u},1}] \in \mathbb{C}^{M_{u} \times \delta_{u}}$, and ${\bm W}_{u}=[{\bm w}_{u,1},\cdots,{\bm w}_{u,\delta_u}]$.
We apply an linear minimum mean square error (LMMSE) estimator to estimate ${\bm Q}_{1,1,u,1}=[{\bm q}_{u,1},\cdots,{\bm q}_{u,M_u}]$ based on $\bar{\bm Y}_{u,1}$:
\vspace{-5pt}
{\fontsize{9pt}{10pt}
\begin{align}\label{es_Q_u1}
    &\hat{\bm Q}_{1,1,u,1}=[\hat{\bm q}_{u,1},\cdots,\hat{\bm q}_{u,M_u}]
    ={\rm unvec}(\alpha^{-1}{\bm C}_q{\bm A}^H({\bm A}{\bm C}_q{\bm A}^H \notag\\
    &+{2\sigma^2}/{|\alpha|^2}{\bm I}_{N{\delta}_u})^{-1}\bar{\bm y}_{u,1}),~~\forall u,
\end{align}}where $\alpha=\sqrt{p}(e^{{\rm j}\theta}-1)$, ${\bm A}={\bm \Psi}_{u,1}^T\otimes {\bm I}_N$, ${\bm C}_q={\bm I}_{M_u}\otimes {\bm C}_{Q,u}$ with
${\bm C}_{Q,u}=\mathbb{E}[{\bm Q}_{1,1,u,1}{\bm Q}_{1,1,u,1}^H]/M_u$ denoting the covariance matrix of ${\bm Q}_{1,1,u,1}$ under the assumption that the columns of ${\bm Q}_{1,1,u,1}$ are i.i.d., and $\bar{\bm y}_{u,1}={\rm vec}(\bar{\bm Y}_{u,1})$. And the MSE is ${\rm tr}({\bm C}_{\Delta q})$, where ${\rm tr}(\cdot)$ denotes the trace of a matrix and
\begin{align}\label{C_Delta_q}
	{\bm C}_{\Delta q}=({|\alpha|^2}/{(2\sigma^2)}{\bm A}^H{\bm A}+{\bm C}_q^{-1})^{-1}.
\end{align}

\begin{remark}\label{remark}
	\textit{Under the group-by-group strategy based on  \cite{wang2025lowoverhead}, it requires $U$ sets of base signals, each set as in \eqref{y_t_base} and $U$ sets of variant signals as in \eqref{y_t_modified}. While in our proposed scheme, it only requires one set of base signals as in \eqref{y_t_base} and $U$ sets of variant signals as in \eqref{y_t_modified}. Therefore, the proposed scheme significantly reduces the overhead to estimate ${\bm Q}_{1,1,u,1}$'s.}
\end{remark}

Having estimated the reference channels ${\bm Q}_{1,1,u,1}$'s, we now proceed to determine the scaling coefficients $\beta_{1,1,u,m}$'s, $m=2,\cdots,M_u$, $\forall u$. After ${\bm Q}_{1,1,u,1}$'s are estimated, the received signals at time instants $t=1,\cdots,\tau_{1,1}$ can be re-written as:
\vspace{-2pt}
\begin{align}\label{y_11}
    {\bm y}_t&={\sum}_{u=1}^{U}{\sum}_{m=2}^{M_u}\beta_{1,1,u,m}{\bm f}_{u,t,m}+{\bm c}_t+{\bm e}_t+{\bm z}_t \notag\\
    &={\bm F}_t\bar{\boldsymbol{\beta}}_{1,1}+{\bm c}_t+{\bm e}_t+{\bm z}_t,~~t=1,\cdots,\tau_{1,1},
\end{align}
where ${\bm c}_t={\sum}_{u=1}^{U}\sqrt{p}a_t\hat{\bm Q}_{1,1,u,1}{\bm \phi}_{u,1,t}$, ${\bm e}_{t}={\sum}_{u=1}^U\sqrt{p}a_t\cdot\\{\bm Q}_{1,1,u,1}{\bm \phi}_{u,1,t}-{\bm c}_t$,  ${\bm F}_t=\sqrt{p}a_t[{\bm Q}_{1,1,1,1},\cdots,{\bm Q}_{1,1,U,1}]\cdot{\rm blkdiag}(\tilde{\bm \Phi}_{1,t},\cdots,\tilde{\bm \Phi}_{U,t})\in\mathbb{C}^{N\times (M-U)}$ with $\tilde{\bm \Phi}_{u,t}=[{\bm \phi}_{u,t,2},\\\cdots,{\bm \phi}_{u,t,M_u}]$, and $\bar{\boldsymbol{\beta}}_{1,1}=[\bar{\bm \beta}_{1,1,1}^T,\cdots,\bar{\bm \beta}_{1,1,U}^T]^T$ with $\bar{\bm \beta}_{1,1,u}=[\beta_{1,1,u,2},\cdots,\beta_{1,1,u,M_u}]^T$, $u=1,\cdots,U$. Then, the overall effective received signal used for estimating $\bar{\bm \beta}_{1,1}$ is defined as
\begin{align}\label{tilde_y}
    \tilde{\bm y}={\bm \Theta}_1\bar{\bm \beta}_{1,1}+\tilde{\bm c}+\tilde{\bm e}+\tilde{\bm z},
\end{align}
where $\tilde{\bm x}=[{\bm x}_1^T,\cdots,{\bm x}_{\tau_{1,1}}^T]^T$ with ${\bm x}\in\{{\bm y},{\bm c},{\bm e},{\bm z}\}$, and ${\bm \Theta}_1=[{\bm F}_1^T,\cdots,{\bm F}_{\tau_{1,1}}^T]^T$. $\tilde{\bm y}$ in \eqref{tilde_y} is re-written as
\begin{align}\label{tilde_y_2}
    \tilde{\bm y}=\hat{\bm \Theta}_1\bar{\bm \beta}_{1,1}+({\bm \Theta}_1-\hat{\bm \Theta}_1)\bar{\bm \beta}_{1,1}+\tilde{\bm c}+\tilde{\bm e}+\tilde{\bm z},
\end{align}
where $\tilde{\bm \Theta}_1$ is as the same form as ${\bm \Theta}_1$ with ${\bm F}_t$ replaced by $\hat{\bm F}_t=\sqrt{p}a_t[\hat{\bm Q}_{1,1,1,1},\cdots,\hat{\bm Q}_{1,1,U,1}]\cdot{\rm blkdiag}(\tilde{\bm \Phi}_{1,t},\cdots,\tilde{\bm \Phi}_{U,t})$, $t=1,\cdots,\tau_{1,1}$. Based on \eqref{tilde_y_2}, the LMMSE estimator of $\bar{\bm \beta}_{1,1}$ is designed as
\begin{align}\label{es_beta11}
	\hat{\boldsymbol{\beta}}_{1,1}
	={\bm C}_{\beta}\hat{\bm \Theta}_1^H(\hat{\bm \Theta}_1{\bm C}_{\beta}\hat{\bm \Theta}_1^H+{\bm C}_{\eta})^{-1}(\tilde{\bm y}-\tilde{\bm c}),
\end{align}
where ${\bm C}_{\beta}$ denotes the covariance matrix of ${\bar{\bm \beta}}_{1,1}$, ${\bm C}_{\eta}={\bm C}_{\Delta \Theta}+{\bm C}_e+ \sigma^2{\bm I}$, ${\bm C}_{\Delta \Theta}$ is the covariance matrix of $({\bm \Theta}_1-\hat{\bm \Theta}_1)\bar{\bm \beta}_{1,1}$ with the $(t,s)$-th sub-block, $t,s=1,\cdots,\tau_{1,1}$, being 
\vspace{-5pt}
{\fontsize{8pt}{12pt}
\begin{align}\label{C_Delta_Theta}
	\mathbb{E}[ (\Delta {\bm F}_t\bar{\bm \beta}_{1,1})(\Delta {\bm F}_s\bar{\bm \beta}_{1,1}) ^H]=pa_ta_s^*\sum_{u=1}^U\sum_{ m=1}^{M_u}\sum_{n=1}^{M_u}{\xi}^{u,t,s}_{m,n}\mathbb{E}[\Delta{\bm q}_{u,m}\Delta{\bm q}_{u,n}^H],
\end{align}}where ${\bm \Xi}^{u,t,s}=\tilde{\bm \Phi}_{u,t}{\bm C}_{\beta,u}\tilde{\bm \Phi}_{u,s}^H$ with $\xi^{u,t,s}_{m,n}$ being its $(m,n)$-th element, $\Delta{\bm F}_t={\bm F}_t-\hat{\bm F}_t$, ${\bm C}_{\beta,u}$ is the covariance matrix of ${\bm \beta}_{1,1,u}$, $\Delta{\bm q}_{u,m}={\bm q}_{u,m}-\hat{\bm q}_{u,m}$, with $\mathbb{E}[\Delta{\bm q}_{u,m}\Delta{\bm q}_{u,n}^H]$ being the ${(m,n)}$-th sub-block of ${\bm C}_{\Delta q}$ in \eqref{C_Delta_q},
and ${\bm C}_e$ is the covariance matrix of $\tilde{\bm e}$ with the $(t,s)$-th sub-block, $t,s=1,\cdots,\tau_{1,1}$, being 
\vspace{-8pt}
\fontsize{9pt}{12pt}{
\begin{align}\label{C_e}
	\mathbb{E}[{\bm e}_t{\bm e}_s^H]=pa_ta_s^*\sum_{u=1}^U\sum_{m=1}^{M_u}\sum_{n=1}^{M_u}{\upsilon}^{u,t,s}_{m,n}\mathbb{E}[\Delta{\bm q}_{u,m}\Delta{\bm q}_{u,n}^H],
\end{align}}where ${\bm \Upsilon}^{u,t,s}={\bm\phi}_{u,t,1}{\bm \phi}_{u,s,1}^H$ with ${\upsilon}^{u,t,s}_{m,n}$ its $(m,n)$-th element. Based on \eqref{es_beta11}, the error covariance matrix of $\hat{\bm \beta}_{1,1}$ is
\begin{align}
	{\bm C}_{\Delta \beta}=(\hat{\bm \Theta}_1^H{\bm C}_{\eta}^{-1}\hat{\bm \Theta}_1+{\bm C}_{\beta}^{-1})^{-1},
\end{align}
and the MSE is ${\rm tr}({\bm C}_{\Delta \beta})$. \eqref{C_Delta_Theta} and \eqref{C_e} show that the error propagated from \eqref{es_Q_u1}, i.e., ${\bm C}_{\Delta \Theta}$ and ${\bm C}_{e}$, scale with ${\bm C}_{\Delta q}$. Moreover, \eqref{C_Delta_q} shows that ${\bm C}_{\Delta q}$ decreases monotonically in the positive semidefinite sense as the SNR $p/\sigma^2$ increases. Consequently, the impact of propagated error terms ${\rm tr}({\bm C}_{\Delta \Theta})$ and ${\rm tr}({\bm C}_{e})$ become negligible at high SNR. Thus, we approximate ${\bm \Theta}_1-\hat{\bm \Theta}_1 \approx {\bm 0}$ and ${\bm e}_t \approx {\bm 0}$ in the remainder of this letter.

\subsection{Phase II: Estimation of the Scaling Coefficients of the Other Antennas of User 1 and the Other Users}

After ${\bm Q}_{1,1,u,1}$'s and $\beta_{1,1,u,m}$'s are estimated in Phase I, the remaining unknown variables for each group $u=1,\cdots,U$ to estimate reduce to
\begin{align}
	\bar{\bm B}_u=[{\bm \beta}_{1,2,u},\cdots,{\bm \beta}_{1,V,u},{\bm B}_{2,u},\cdots,{\bm B}_{K,u}].	
\end{align}
In the following, we introduce how to estimate $\bar{\bm B}_u$'s in Phase II. Specifically, in Phase II at time instants $t=\tau_1+1,\cdots,\tau_1+\tau_2$, we keep the $1$st antenna of user $1$ silent, i.e., $a_{1,1,t}=0,t=\tau_1+1,\cdots,\tau_1+\tau_2$, while for the other antennas of user $1$ and all antennas of users $2$ to $K$, we define the overall transmit pilot symbols as $\bar{\bm a}_t=[a_{1,2,t},\cdots,a_{1,V,t},{\bm a}_{2,t}^T,\cdots,{\bm a}_{K,t}^T]^T\in\mathbb{C}^{(KV-1)\times 1}$. Then, the received signals in \eqref{y_t_new} over $\tau_2$ time instants can be re-written as
\begin{align}\label{y_t_P2}
	{\bm y}_t&={\sum}_{u=1}^U{\bm Q}_{1,1,u,1}{\bm \Phi}_{u,t}\bar{\bm B}_u\sqrt{p}\bar{\bm a}_{t}+{\bm z}_t={\sum}_{u=1}^U{\bm P}_{u,t}\bar{\bm b}_u+{\bm z}_t\notag\\
	&={\bm T}_t\bar{\bm b}+{\bm z}_t,~~t=\tau_1+1,\cdots,\tau_1+\tau_2,
\end{align}
where ${\bm T}_t=[{\bm P}_{1,t},\cdots,{\bm P}_{U,t}]\in\mathbb{C}^{N\times M(KV-1)}$ with ${\bm P}_{u,t}=\sqrt{p}\bar{\bm a}_{t}^T\otimes{\bm Q}_{1,1,u,1}{\bm \Phi}_{u,t}\in\mathbb{C}^{N\times M_u(KV-1)}$, and $\bar{\bm b}=[\bar{\bm b}_1^T,\cdots,\bar{\bm b}_U^T]^T$ with $\bar{\bm b}_u={\rm vec}(\bar{\bm B}_u)$. The overall received signal of the BS over Phase II is given by
\begin{align}\label{y^2}
	{\bm y}^{(2)}&=[{\bm y}_{\tau_1+1}^T,\cdots,{\bm y}_{\tau_1+\tau_2}^T]^T={\bm \Theta}_2\bar{\bm b}+{\bm z}^{(2)},
\end{align}
where ${\bm \Theta}_2=\left[{\bm T}_{\tau_1+1}^T,\cdots,{\bm T}_{\tau_1+\tau_2}^T	\right]^T\in\mathbb{C}^{N\tau_2\times M(KV-1)}$ and ${\bm z}^{(2)}=[{\bm z}_{\tau_1+1}^T,\cdots,{\bm z}_{\tau_1+\tau_2}^T]^T$. ${\bm y}^{(2)}$ can be re-expressed as
\begin{align}\label{y^2+n}
	{\bm y}^{(2)}=\hat{\bm \Theta}_2\bar{\bm b}+({\bm \Theta}_2-\hat{\bm \Theta}_2)\bar{\bm b}+{\bm z}^{(2)},
\end{align}
where $\hat{\bm \Theta}_2$ has the same form as ${\bm \Theta}_2$ with ${\bm Q}_{1,1,u,1}$'s replaced by $\hat{\bm Q}_{1,1,u,1}$'s given in \eqref{es_Q_u1}. Similar to Phase I, we assume that the error propagated from \eqref{es_Q_u1}, i.e., ${\bm \Theta}_2-\hat{\bm \Theta}_2$, can be reduced as much as possible. In this case, \eqref{y^2+n} is approximated to
\begin{align}\label{y^2_approx}
	{\bm y}^{(2)}\approx\hat{\bm\Theta}_2\bar{\bm b}+{\bm z}^{(2)}.
\end{align}
We generate $\bar{\bm a}_{t}$'s as random vectors, and ${\bm \Phi}_{u,t}$'s as random unitary matrices. Then, based on \eqref{y^2_approx}, the LMMSE estimator of $\bar{\bm b}$ can be designed as
\begin{align}\label{es_beta}
	\hat{{\bm b}}={\bm C}_{b}\hat{\bm \Theta}_2^H(\hat{\bm \Theta}_2{\bm C}_{b}\hat{\bm \Theta}_2^H+\sigma^2{\bm I}_{N\tau_2})^{-1}{\bm y}^{(2)},
\end{align}
where ${\bm C}_{b}$ denotes the covariance matrix of ${\bar{\bm b}}$. Finally, the overall cascaded channels $\hat{\bm Q}_{k,v,u,m}$ 's are reconstructed using the estimated $\hat{\bm Q}_{1,1,u,1}$'s by \eqref{es_Q_u1} and $\hat{\beta}_{k,v,u,m}$'s by \eqref{es_beta11} and \eqref{es_beta} via \eqref{corr_u}: $\hat{\bm Q}_{k,v,u,m}=\hat{\beta}_{k,v,u,m}\hat{\bm Q}_{1,1,u,1}$, $\forall u, (k,v,m)\neq(1,1,1)$, thus constituting estimated $\hat{\bm J}_{k,u}$'s based on \eqref{Jku}.

\begin{table}[t]
	\begingroup
	\renewcommand{\arraystretch}{1.2}
	\newcolumntype{C}{>{\centering\arraybackslash}X}
	\centering
	\caption{Comparison of Channel Estimation Overhead}\label{tab1}
	
	\begin{tabularx}{\columnwidth}{|>{\centering\arraybackslash}m{0.4\columnwidth}|C|}
		\hline
		Proposed Scheme & less than $2M+U\lceil[\bar{M}(KV-1)]/q\rceil$ \\
		\hline
		Benchmark Scheme I and II & $KVU\bar{M}^2$ \\
		\hline
		Benchmark Scheme III& less than $KVU\bar{M}^2$\\
		\hline
		Benchmark Scheme IV & $2M+U\lceil[\bar{M}(KV-1)]/q\rceil$ \\
		\hline
	\end{tabularx}
	\endgroup
	\vspace{-15pt}
\end{table}

\section{Numerical Results}\label{simulation}

In this section, we provide numerical examples to demonstrate the advantages of our proposed scheme. The BS-RIS channel ${\bm G}$ follows a Rician fading model $\textstyle {\bm G}=\sqrt{{\kappa}/{(1+\kappa)}}\bar{\bm G}+\sqrt{{1}/{(1+\kappa)}}({\bm C}^{\rm B})^{\frac{1}{2}}{\bm G}_w({\bm C}^{\rm R})^{\frac{1}{2}}$,
where $\kappa$ denotes the Rician factor set as $10$dB, $\bar{\bm G}$ denotes the LoS component, ${\bm C}^{\rm B}$ and ${\bm C}^{\rm R}$ denote the BS 
correlation matrix and the RIS correlation matrix, respectively, generated based on the exponential correlation matrix model \cite{wang2020channel}, and ${\bm G}_w\sim\mathcal{CN}({\bm 0},M\ell^{\rm BR}{\bm I})$ denotes the
i.i.d. Rayleigh fading component with
$\ell^{\rm BR}$ being the pass loss of ${\bm G}$. The channel between the $v$-th antenna of user $k$ and the $m$-th BD-RIS element in group $u$ follows $r_{k,v,u,m}\sim\mathcal{CN}(0,\ell^{\rm UR})$, where $\ell^{\rm UR}$ denotes the path loss \cite{wang2025lowoverhead}. The transmit power of the user is $p=33$ dBm. The power spectrum density of the noise at the BS is assumed to be $-169$ dBm/Hz, and the channel bandwidth is $1$ MHz. The normalized mean-squared error (NMSE) is used as the metric to evaluate the performance of channel estimation, which is defined as
$
	{\rm NMSE}=\mathbb{E}\left[\frac{1}{K}{\sum}_{k=1}^K\frac{||\hat{\bm J}_{k}-{\bm J}_{k}||_F^2}{||{\bm J}_{k}||_F^2}\right],
$
where $\hat{\bm J}_k=[\hat{\bm J}_{k,1},\cdots,\hat{\bm J}_{k,U}]$, $\forall k$.

To show the performance gain of our proposed scheme, we adopt the following benchmark schemes, which directly estimate all entries in ${\bm J}_{u}$'s based on \eqref{rev}:
\textbf{Benchmark Scheme I}: The LS estimator proposed in \cite{li2024channel}; \textbf{Benchmark Scheme II}: The Block Tucker Kronecker factorization (BTKF) algorithm \cite{de2024channel}; \textbf{Benchmark Scheme III}: The Block Tucker alternating least squares (BTALS) algorithm\cite{de2024channel}; \textbf{Benchmark Scheme IV}: The group by group scheme based on \cite{wang2025lowoverhead}. Assuming each BD-RIS group has the same size $\bar{M}=M/U$ and $q$ is the rank of ${\bm G}_u$, a  comparison between the theoretical overhead is shown in Table \ref{tab1}.
\begin{figure}
	\vspace{2pt}
	\centering
	\begin{subfigure}[t]{0.21\textwidth}
		\centering
		\includegraphics[width=1.1\textwidth]{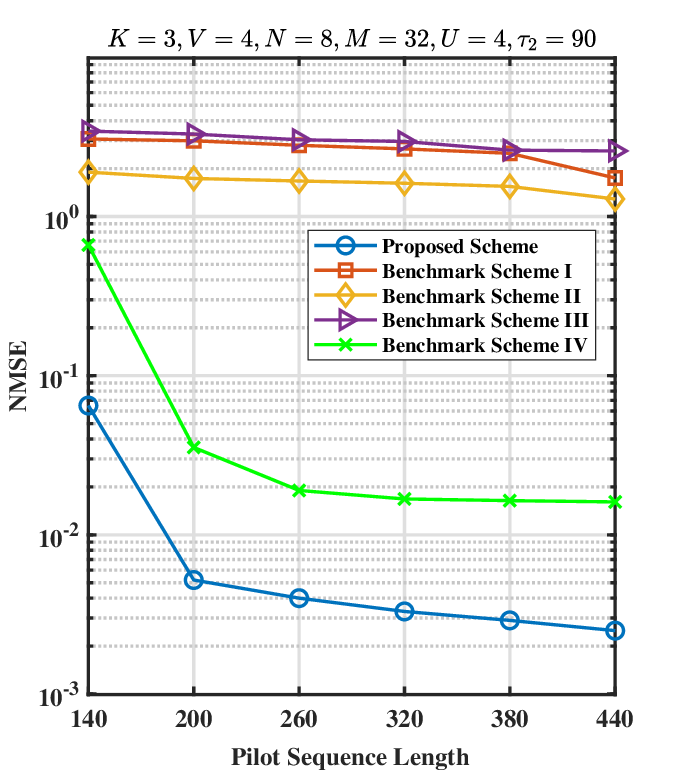}\vspace{-5pt}
		\caption{\footnotesize NMSE versus $\tau$ when $M=32,N=8,U=4,K=3,V=4,\tau_2=90$ and random ${\bm \Phi}_t$'s.}
	\end{subfigure}
	~
	\begin{subfigure}[t]{0.21\textwidth}
		\centering
		\includegraphics[width=1.1\textwidth]{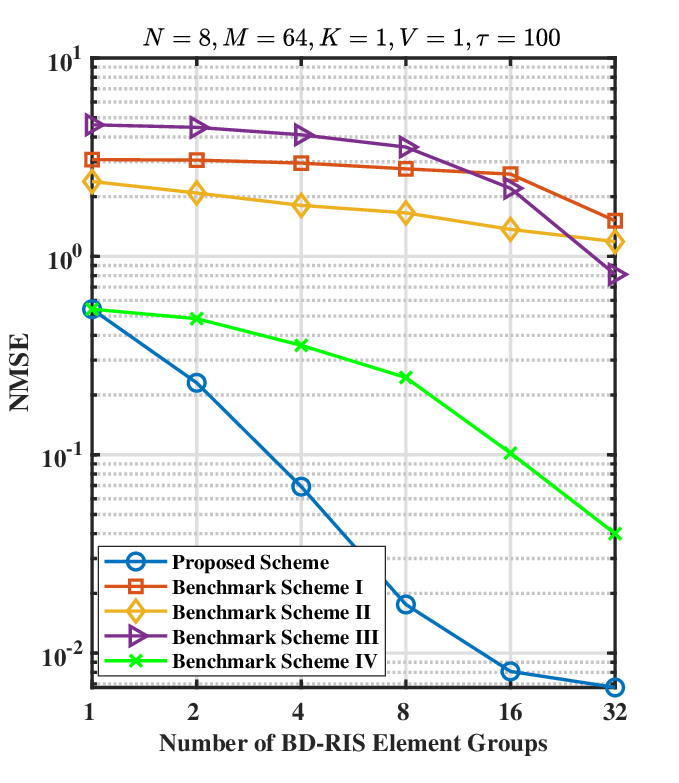}\vspace{-5pt}
		\caption{\footnotesize NMSE versus $U$ when $M=64,N=8,K=1,V=1,\tau=100$ and random ${\bm \Phi}_t$'s .}
	\end{subfigure}
	\caption{NMSE performance comparison between the proposed scheme and benchmark schemes.}
	\label{fig2}
	\vspace{-10pt}
\end{figure}

Fig. \ref{fig2} shows the NMSE performance comparison between our proposed scheme and the benchmark schemes. In Fig 2(a), the number of BS antennas, BD-RIS elements, users, user antennas, and BD-RIS groups are set as $N=8$, $M=32$, $K=3$,  $V=4$ and $U=4$, respectively. Each group has $\bar{M}=8$ elements. The total pilot sequence length ranges from $140$ to $440$. It is observed that our proposed scheme shows a significant NMSE performance gain, because Benchmark Schemes I to III treat all the channel coefficients as independent variables, while Benchmark Scheme IV uses redundant pilot signals to estimate reference channels as mentioned in Remark \ref{remark}. Furthermore, Fig. 2(b) shows the NMSE performance versus the number of BD-RIS elements groups where $N=8$, $K=1$, $V=1$, and $M=64$. The number of groups $U$ ranges among $1,2,4,8,16,32$. The NMSE of our proposed scheme decreases with $U$ increasing, because more pilot signals can be saved to estimate reference channels.

\vspace{-5pt}
\section{Conclusions}\label{conclude}

This letter addressed channel estimation for a multi-user MIMO communication system assisted by a group-connected BD-RIS architecture. By exploiting a fundamental property that cascaded channels are scaled versions of a reference channel associated with one element within a group, we proposed a two-phase channel estimation protocol, where the reference cascaded channel matrices and the scaling coefficients are sequentially estimated. Numerical results verified that our proposed method accurately estimates channels with substantially lower overhead than existing methods.

\bibliographystyle{IEEEtran}
\bibliography{reference}

@ARTICLE{shen2022Modeling,
  author={Shen, Shanpu and Clerckx, Bruno and Murch, Ross},
  journal={IEEE Trans. Wireless Commun.}, 
  title={Modeling and Architecture Design of Reconfigurable Intelligent Surfaces Using Scattering Parameter Network Analysis}, 
  year={Feb. 2022},
  volume={21},
  number={2},
  pages={1229-1243},
}

@ARTICLE{wang2020channel,
  author={Wang, Zhaorui and Liu, Liang and Cui, Shuguang},
  journal={IEEE Trans. Wireless Commun.}, 
  title={Channel Estimation for Intelligent Reflecting Surface Assisted Multiuser Communications: Framework, Algorithms, and Analysis}, 
  year={Oct. 2020},
  volume={19},
  number={10},
  pages={6607-6620}}

@ARTICLE{li2024channel,
  author={Li, Hongyu and Shen, Shanpu and Zhang, Yumeng and Clerckx, Bruno},
  journal={IEEE Trans. Signal Process.}, 
  title={Channel Estimation and Beamforming for Beyond Diagonal Reconfigurable Intelligent Surfaces}, 
  year={Jul. 2024},
  volume={72},
  number={},
  pages={3318-3332},
}

@ARTICLE{Li2023BDRIS,
  author={Li, Hongyu and Shen, Shanpu and Nerini, Matteo and Clerckx, Bruno},
  journal={IEEE Commun. Mag.}, 
  title={Reconfigurable Intelligent Surfaces 2.0: Beyond Diagonal Phase Shift Matrices}, 
  year={Mar. 2024},
  volume={62},
  number={3},
  pages={102-108}}

@ARTICLE{de2024channel,
	author={de Almeida, Andre L. F. and Sokal, Bruno and Li, Hongyu and Clerckx, Bruno},
	journal={IEEE Trans. Signal Process.}, 
	title={Channel Estimation for Beyond Diagonal {RIS} via Tensor Decomposition}, 
	year={early access, May 2025}}

@ARTICLE{ginige2024prediction,
  author={Ginige, Nipuni and de Sena, Arthur Sousa and Mahmood, Nurul Huda and Rajatheva, Nandana and Latva-Aho, Matti},
  journal={IEEE Trans. Veh. Tech.}, 
  title={Efficient Channel Prediction for Beyond Diagonal {RIS}-Assisted {MIMO} Systems With Channel Aging}, 
  year={Mar. 2025},
  volume={74},
  number={8},
  pages={12658-12672}
}

@INPROCEEDINGS{dearaujo2024semiblind,
  author={de Araújo, Gilderlan Tavares and de Almeida, André L. F.},
  booktitle={Asilomar Conf. Signals, Syst., and Comput.}, 
  title={Semi-Blind Channel Estimation for Beyond Diagonal {RIS}}, 
  year={Dec. 2024},
  volume={},
  number={},
  pages={1586-1590}
}

@ARTICLE{Basar2019wireless,
  author={Basar, Ertugrul and Di Renzo, Marco and De Rosny, Julien and Debbah, Merouane and Alouini, Mohamed-Slim and Zhang, Rui},
  journal={IEEE Access},
  title={Wireless communications through reconfigurable intelligent surfaces},
  year={Aug. 2019},
  volume={7},
  number={},
  pages={116 753--116 773}
}

@ARTICLE{jian2022inte,
  author={Jian, Mengnan and Alexandropoulos, George C. and Basar, Ertugrul and Huang, Chongwen and Liu, Ruiqi and Liu, Yuanwei and Yuen, Chau},
  journal={Intell. Conv. Networks}, 
  title={Reconfigurable intelligent surfaces for wireless communications: Overview of hardware designs, channel models, and estimation techniques}, 
  year={Mar. 2022},
  volume={3},
  number={1},
}

@ARTICLE{wang2025lowoverhead,
        author={Wang, Rui and Zhang, Shuowen and Clerckx, Bruno and Liu, Liang},
      journal={IEEE Trans. Signal Process.}, 
      title={Low-Overhead Channel Estimation Framework for Beyond Diagonal Reconfigurable Intelligent Surface Assisted Multi-User {MIMO} Communication}, 
      year={Nov. 2025},
      volume={73},
      number={},
      pages={4700-4717},
}

@ARTICLE{chen2023channel,
	author={Chen, Jie and Liang, Ying-Chang and Cheng, Hei Victor and Yu, Wei},
	journal={IEEE Trans. Wireless Commun.}, 
	title={Channel Estimation for Reconfigurable Intelligent Surface Aided Multi-User mm{W}ave {MIMO} Systems}, 
	year={Oct. 2023},
	volume={22},
	number={10},
	pages={6853-6869}}

\end{document}